\documentclass{aa} 
 
\usepackage{graphicx} 
\usepackage{natbib} 
\usepackage{txfonts} 
 
\begin{document} 
 
\title {Chemical evolution models for spiral disks: the Milky Way, M31 and M33  } 
 
\author {M. M. Marcon-Uchida\inst{1,2} \thanks {email to: marcon@oats.inaf.it /  monica@astro.iag.usp.br} 
  \and F. Matteucci\inst{1,3} \and R. D. D. Costa\inst{2} \institute{ 
  Dipartimento di Fisica, Sezione di Astronomia, Universit\`a di Trieste, via 
  G.B. Tiepolo 11, I-34131, Trieste, Italy \and Instituto de Astronomia, 
  Geof\'\i sica e Ci\^encias Atmosf\'ericas (IAG), Universidade de S\~ao Paulo, 
  Rua do Mat\~ao, 1226 Cidade Universit\'aria, S\~ao Paulo - SP, 05508-900, Brazil 
  \and I.N.A.F. Osservatorio Astronomico di 
  Trieste, via G.B. Tiepolo 11, I-34131, Trieste, Italy}} 
 
\date{Received xxxx / Accepted xxxx} 
 
\abstract{The distribution of chemical abundances and their variation in  
time are important tools to understand the chemical evolution of galaxies: in particular,  
the study of chemical evolution models 
can improve our understanding of the basic assumptions made for modelling our Galaxy and other spirals.} 
{To test a standard chemical evolution model for spiral disks in the Local Universe and study the  
influence of a 
threshold gas density and different efficiencies in the star formation rate (SFR)  
law on radial gradients (abundance, gas and SFR). The model will be then applied to specific galaxies.} 
{We adopt a one-infall chemical evolution model where the Galactic disk forms inside-out by means of 
infall of gas, and 
we test different thresholds and efficiencies in the SFR. The model is scaled to the disk 
properties of three Local Group galaxies (the Milky Way, M31 and M33) by varying its dependence on   
the star formation efficiency and the time scale for the infalling gas into the disk. 
} 
{Using this simple model we are able to reproduce most of the observed constraints available in the literature for the studied galaxies. The  
radial oxygen abundance gradients and their time evolution are studied in detail. The present day abundance gradients 
 are more sensitive to the threshold  than to other parameters, while their  temporal evolutions are more dependent 
on the chosen SFR efficiency. A variable efficiency along the galaxy radius can reproduce the present day gas distribution 
in the disk of spirals with prominent arms. The steepness in the distribution of stellar surface density differs from 
massive to lower mass disks, owing to the different star formation histories.  
 } 
{The most massive disks seem to have evolved faster  (i.e. with more efficient star formation) than the less massive ones, thus suggesting a downsizing in star formation for spirals. The threshold and the efficiency of star formation play a very important role in the chemical evolution of spiral disks and 
an efficiency varying with radius can be used to regulate the star formation. The oxygen abundance gradient can steepen or flatten in time depending on the choice of this parameter.}
 
% \abstract{.} 
 

\titlerunning{Chemical evolution of the MW, M31 and M33} 
\authorrunning{Marcon-Uchida et al.} 
\maketitle 
  
\section{Introduction} 
 
%The Local Group of galaxies is a perfect scenary to test chemical evolution models and its basic assumptions. During the last years 
%many observational studies have been made to investigate the chemical and dynamical properties of these neighbours systems. The new 
%generation telescopes have contributed to the exploration of diverse stellar populations and provided more accurate data to the 
%observational constraints for the chemical evolution models.  
 
The study of the chemical evolution of nearby spiral galaxies is very important to improve our knowledge about the main ingredients used 
in chemical evolution models and to test the basic assumptions made for modelling our Galaxy. M31 and M33 are other spiral members  
of the Local Group of galaxies and during recent years many observational studies have been made to  
investigate the chemical and  
dynamical properties of these neighbouring systems. New surveys (Braun et al. 2009, Magrini et al. 2007,2008) contributed to the analysis of different  stellar populations and  
provided more accurate data to constrain  the chemical evolution models.  
 
The disks of M31 and M33 have many similarities with the Milky Way disk but some observational constraints like the present day  
gas distribution can only be explained by assuming different star formation  
histories for these galaxies. The SFR 
is one of the most important parameters regulating the chemical evolution  
of galaxies (Kennicutt 1998, Matteucci 2001, Boissier et al. 2003) together with the initial  
mass function (IMF).

Another important mechanism is the "inside-out" disk formation that is very  
important to reproduce the radial abundance 
gradients (see Colavitti et al. 2008 for the most recent paper on the subject). 
A faster formation of the inner disk  
relative to the outer disk was originally proposed by Matteucci \& Fran\c cois (1989)and  
supported in the following years by Boissier \& Prantzos (1999) and Chiappini  
et al. (2001).

The chemical evolution of M31 in comparison with that of the Milky Way has been already discussed by Renda et al. (2005) and Yin et al. (2009). 
Renda et al. (2005) concluded that while the evolution of the MW and M31 share similar properties, differences in the formation history of these two galaxies are required to explain the observations in detail. In particular, they found that the observed higher metallicity in the M31 halo can be explained by either (i) a higher halo star formation efficiency, or (ii) a larger reservoir of infalling halo gas with a longer halo formation phase. These two different pictures would lead to (a) a higher [O/Fe] at low metallicities, or (b) younger stellar populations in the M31 halo, respectively. Both pictures result in a more massive stellar halo in M31, which suggests  a possible correlation between the halo metallicity and its stellar mass. Yin et al. (2009) concluded that M31 must have been more active in the past than the Milky Way although its current SFR is lower than in the Milky Way, and that our Galaxy must be a rather quiescent galaxy, atypical of its class (see also Hammer et al. 2007). They also concluded that the star formation efficiency in M31 must have been higher by a factor of two than in the Galaxy. However, by adopting the same SFR as in the Milky Way they failed in reproducing the observed radial profile of the star formation and of the gas, and suggested that possible dynamical interactions could explain these distributions. 
 
% Some authors (e.g. Diaz \& Tosi 1984, Moll\`a et al. 1996, 1997 and Magrini et al. 2007) have carried-out 
Magrini  et al. (2007) computed the chemical evolution of the disk of M33: they claimed to reproduce the observational  
features by assuming a continuous almost constant infall of gas. 
 
In this work we present a one-infall chemical evolution model for the Galactic disk based on an updated  
version of the Chiappini et al. (2001) model.  This model can predict the evolution of the abundances of 37 chemical elements from the light to the heavy ones. We use this model to reproduce the chemical evolution of the  
Milky Way disk and that of the two nearby spiral galaxies (M31 and M33). To do that,  we assume that the disk  
of each galaxy  formed by gas accretion and vary the star formation efficiency as well as the gas accretion  
timescale. The similarities and the differences between the chemical evolution of these objects and the  
Milky Way are discussed to provide a basis for the understanding of the chemical evolution of disks.  
 
The paper is organized as follows: in section 2 we describe our  
chemical evolution model and the assumptions made for  
each galaxy. In section 3 we present the results for the models  
and these results are discussed in detail in section 4. 
Finally in section 5 we summarize our conclusions.

\section{The chemical evolution model} 
 
In order to reproduce the chemical evolution of the thin-disk, we adopted an updated one-infall version of the  
chemical evolution 
model presented by Chiappini et al. (2001) (hereafter CMR2001). In this model, the galactic disk is divided into  
several concentric rings which evolve independently without exchange of matter. 
 
The disk is built up in an "inside-out" scenario which is a necessary condition to reproduce the radial  
abundance gradients (Colavitti et al. 2008).  
The infall law for the thin-disk is defined as: 
 
\begin{equation}\label{infall} 
\frac{d\Sigma_I(R,t)}{dt} = B(R)e^{-\frac{(t-t_{max})}{\tau_D}} 
\end{equation} 
 
where $\Sigma_I(R,t)$ is the gas surface density of the infall, $t_{max}$ is the time of maximum gas accretion  
in the disk, set equal to 1 Gyr, coincident with the end of halo 
/thick disk phase and $\tau_D$ is the timescale  
for the infalling gas into the thin-disk. To have an 
inside-out formation in the disk, the timescale for the mass accretion is assumed to increase with the Galactic  
radius following a simple linear relation. In particular, we tested different linear relations, as we will see in  
table \ref{timescale}. The coefficient B(R) is derived from the condition that the total mass surface  
density at the present time in the disk is reproduced.   
 
In order to  make the program as simple and generalized as possible, we used a SFR 
proportional to a Schmidt law: 
 
\begin{equation} 
\Psi(r,t) \propto \nu \Sigma_{gas}^k(r,t) 
\end{equation}

where $\nu$ is the efficiency in the star formation process and the surface gas density is represented by 
$\Sigma_{gas}(r,t)$ while the exponent $k$ is  
equal to 1.5 (see Kennicutt 1998 and Chiappini et al. 1997).
We also assume a threshold in the surface gas  
density for star formation. The existence of such a threshold for the star formation  
in the disk of spiral galaxies  
is still the subject of discussion. It has been proposed by many authors and had its first 
observational evidence in Kennicutt (1998) who noticed that below a lower limit surface  
gas density the  
star formation is suppressed.  According to Colavitti et al (2008) this lower limit in the  
gas surface density  
to regulate the star formation is an important key to reproduce the slope of the abundance gradients in the the  
outer disk of the Milky Way.  In this model, we adopted two values for the threshold in the star formation :  
$4$ and $7 M_{\odot}pc^{-2}$ . This is justified by the fact that the threshold is a very uncertain quantity which can vary from galaxy to galaxy and even inside the same galaxy (Kennicutt 1989).
 
\begin{figure}[ht!] 
\includegraphics[width=0.45\textwidth]{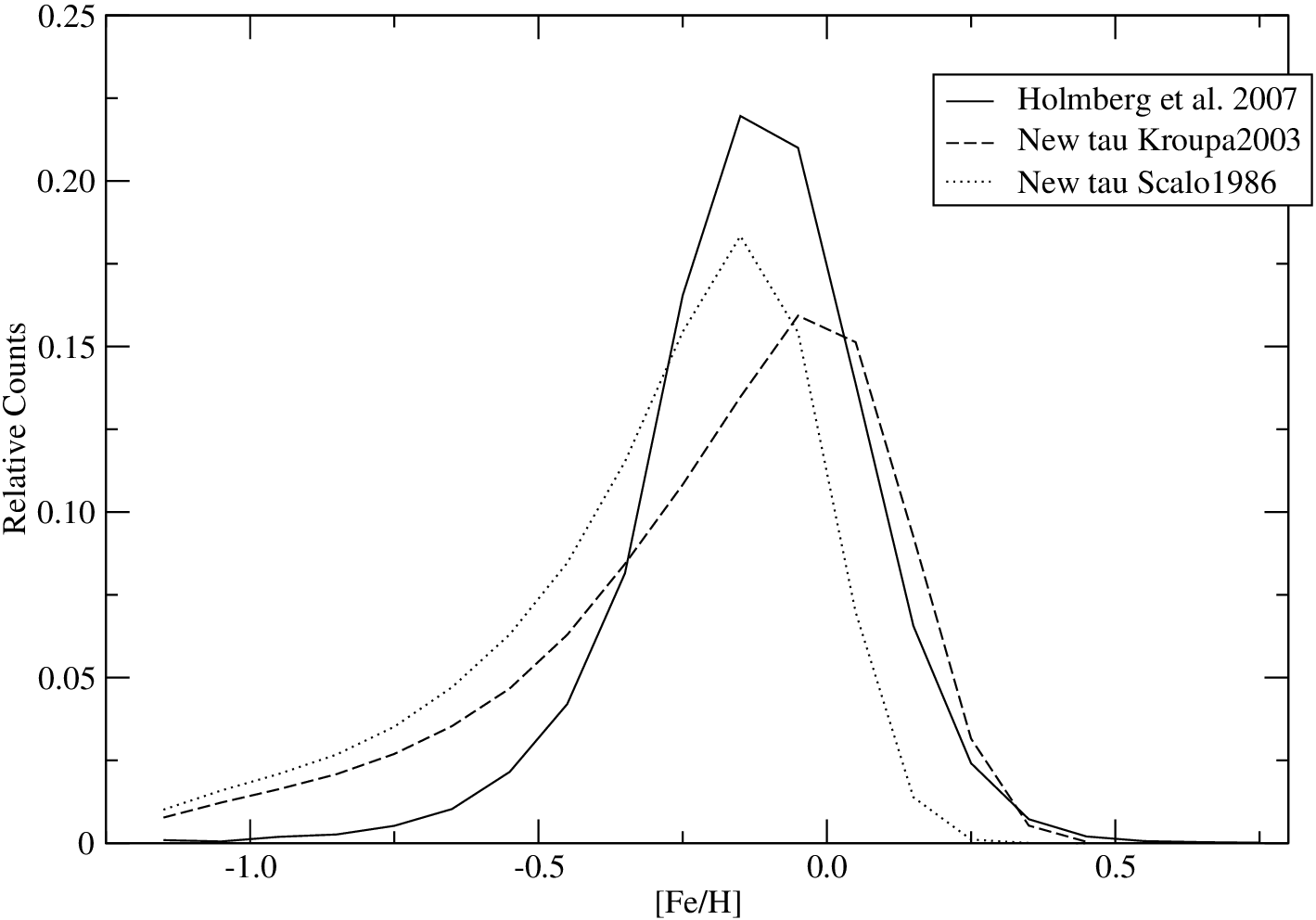} 
\caption{Distribution of dwarf stars in the solar vicinity obtained by using different IMFs. Scalo 1986 (dotted line) and Kroupa et al. (1993) (dashed line) compared to the 
observational data from Holmberg et al. 2007 (solid line). The label ``New tau'' indicates that we have used the $\tau(R)$ law of this paper shown in Table 1.} 
\label{scalo86} 
\end{figure}

%\begin{equation}\label{tau} 
%\tau_D(R) = a R + b 
%\end{equation} 
 
According to recent studies (e.g. Romano et al. 2005) the IMF and the stellar lifetimes are responsable for the  
uncertanties in the 
chemical evolution models for the Milky Way. In this work we assumed an IMF constant in space and time and  
adopted the prescription from  
Kroupa (1993), instead of a two-slope approximation of Scalo (1986) used by CMR2001.   
The total surface mass density distribution for the Galactic disk was assumed to be exponential with scale length $R_D = 3.5$ kpc  
normalized to $\Sigma_D(R_{\odot},t_{Gal}) = 54 M_{\odot}pc^{-2}$ (Romano et al. 2000) 
 
\begin{equation} 
\Sigma_D(R,t_{Gal})= \Sigma_0(0,t_{Gal})e^{-R/R_D} 
\end{equation} 
 
Apart from the IMF, this model differs from the one of the CMR2001 model in: (1) the oxygen yields for  
massive stars that 
are supposed to be metallicity-dependent and  taken from Woosley \& Weaver (1995), as suggested in Fran\c cois et al. (2004); 
(2) the stellar lifetimes of Schaller et al (1992) instead of the Maeder \& Meynet (1989); 
and (3) the solar abundances are those from Asplund et al. (2009). 
%%%%%%%%%%%%%%%%%%%%%%%%%%%%%%%%%%%%%%%%%%%%%%%%%%%%%%%%%%%%%%%%%%%%%%%%%%%%%%%%%%%%%%%%%%%%%%%%%%%%% 
 
\subsection{The Milky Way}

We  computed the model for the Milky Way several times with star formation efficiency of $1Gyr^{-1}$ and different time scales for the infalling  
gas into the disk ($\tau$). Table \ref{timescale} shows the coefficients for the linear equations adopted for $\tau(R)$. 
Figure \ref{scalo86}  
contains  the predictions for the 
dwarf metallicity distribution in the solar neighbourhood (Scalo (1986) represented as a dotted line and Kroupa  et al. (1993)  
represented as a dashed line) 
and the observed corrected values from the Geneva-Copenhagen Survey (GCS) as presented in Holmberg et al. (2007).  
The predictions from the model using the IMF of Kroupa 
(1993) fit better the wings of the dwarf distribution especially for higher metallicities and  produce fewer  
metal-poor stars, reducing  
the well-known "G-dwarf" problem.

\begin{table}[ht!] 
\caption{Coefficients for the timescale equation} 
\label{timescale} 
\begin{center} 
\begin{tabular}{ccc} 
  \hline\hline 
  
\noalign{\smallskip} 
Reference & angular & linear \\ 
 & coef. & coef.\\ 
\noalign{\smallskip} 
 
  \hline 
\noalign{\smallskip} 
 this work & 0.75 & 1.08\\ 
% &  &  \\ 
 CMR2001 & 1.03 & -1.27\\ 
% &  &  \\ 
 Carigi et al. 2008 & 1.0 & -2.0\\ 
% &  &  \\ 
 Renda et al. 2005 & 1.25 & 2.0\\ 
% &  &  \\  
 Boissier \& Prantzos 1999 & 0.86 & 0.14\\ 
 Chang et al. 1999 & 0.30 & 1.38\\
 
\hline 
\end{tabular} 
\end{center} 
\end{table} 
 
\begin{figure}[ht!] 
\includegraphics[width=0.45\textwidth]{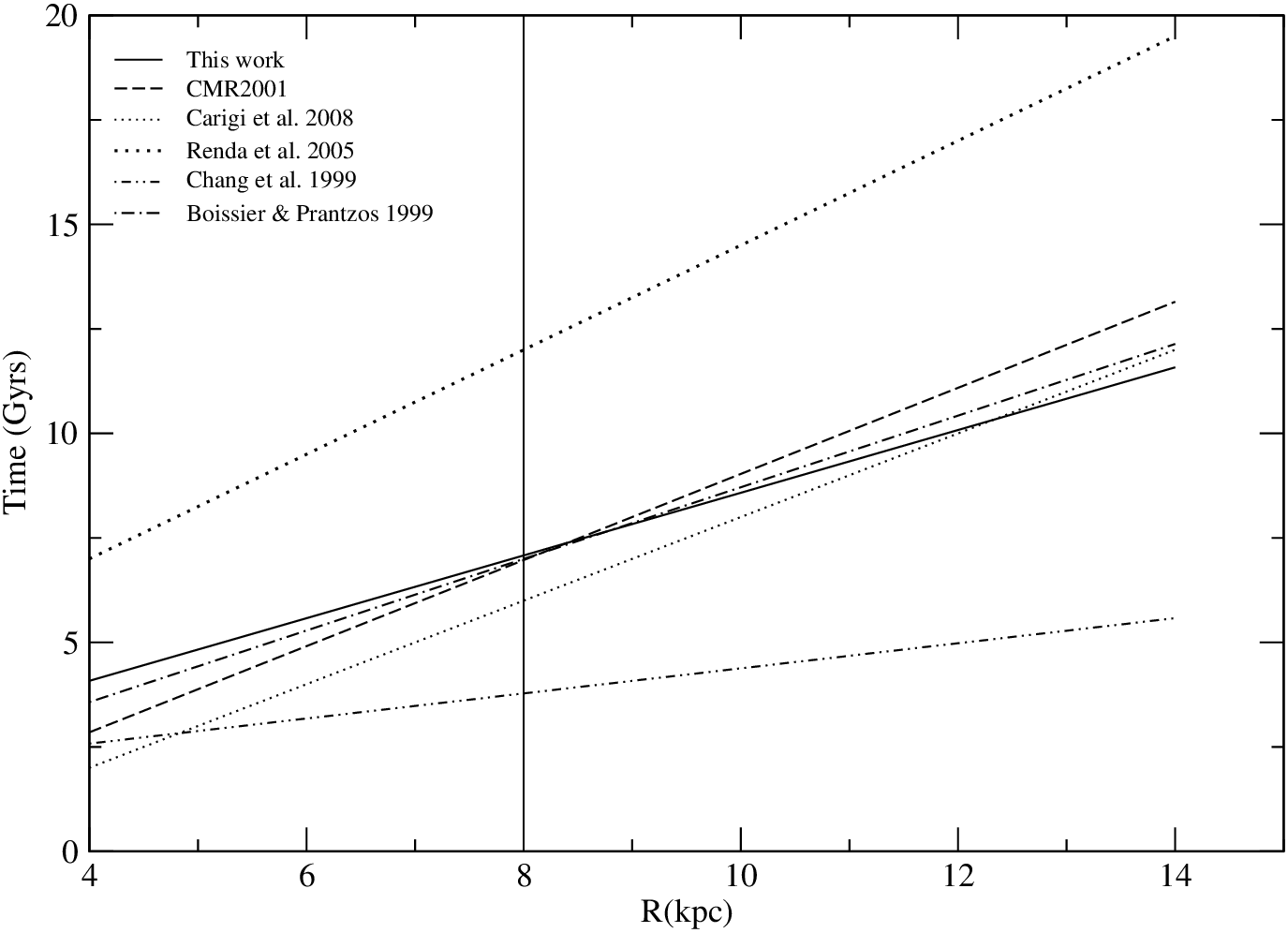} 
\caption{Different time scales for infalling gas along the disk tested in this work. Solid line for this work, dashed line for CMR2001, 
dotted line for Carigi et al. (2008), dot-dashed for Boissier \& Prantzos (1999), filled circles for Renda et al. (2005) and dot-dot-dashed lines for Chang et al. (1999).} 
\label{tau} 
\end{figure}

\begin{figure*}[ht!] 
\includegraphics[width=0.9\textwidth]{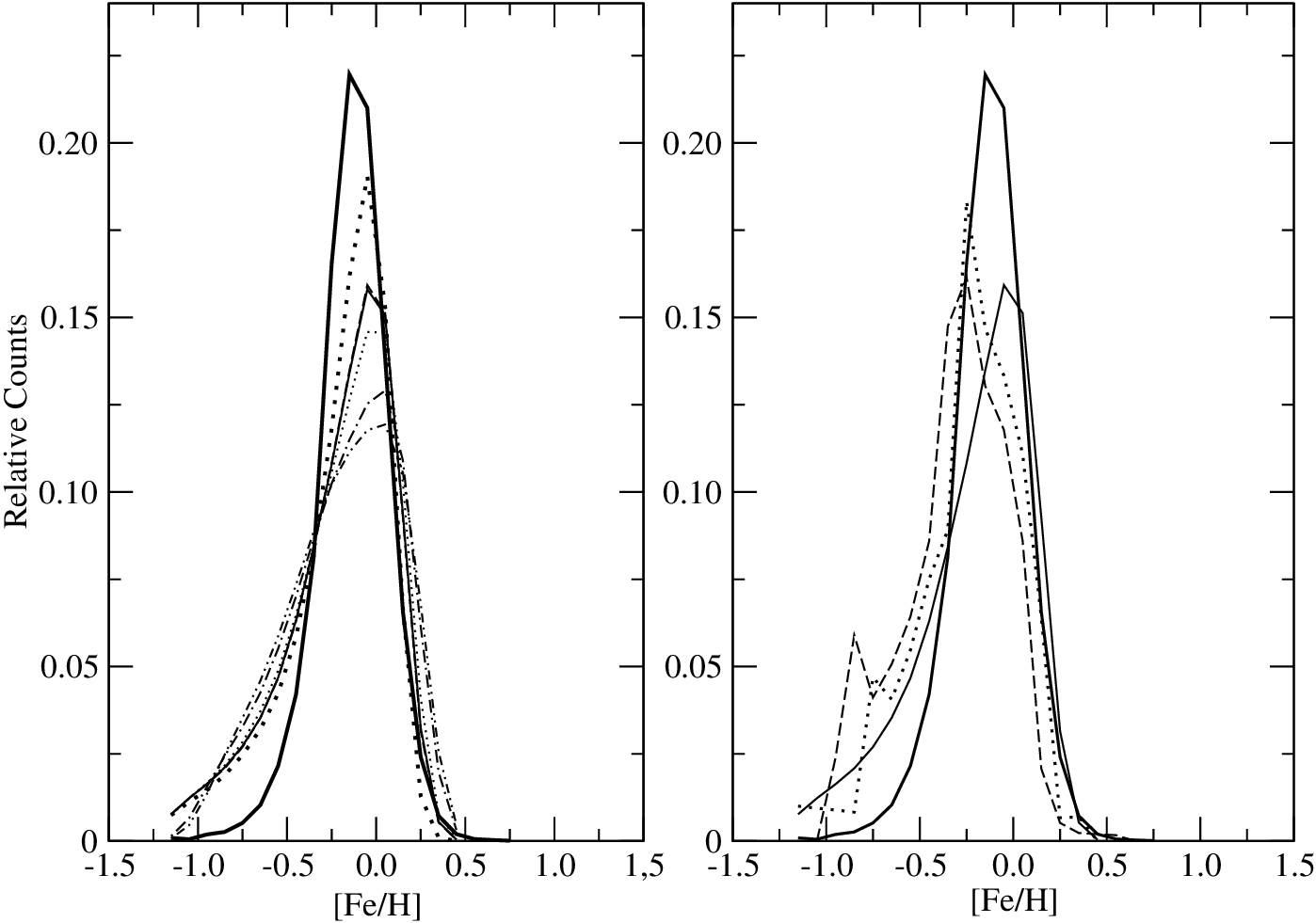} 
\caption{Dwarf star metallicity distributions for the Milky Way. Left panel shows different distributions estimated with the various $\tau$. Solid line for this work, dashed line for CMR2001, 
dotted line for Carigi et al. (2008), dot-dashed for Boissier \& Prantzos (1999), filled circles for Renda et al. (2005) and dot-dot-dashed lines for Chang et al. (1999). Right panel shows a comparison between our model (labelled this work in Table 1) and other dwarf distributions: solid thin line is the   model prediction, solid thick line represents Holmberg et al. (2007) distribution while the dotted one and the dashed line are the  distributions of Kotoneva (2002) and Rocha-Pinto \& Maciel (1996), respectively. } 
\label{nane_holmb} 
\end{figure*}

In Figure \ref{tau} we show the different gas infall laws  tested in this work and taken from the literature; the vertical line marks 
the radius correspondent to the solar vicinity at 8 kpc. Comparing the results of the dwarf metallicity distributions (figure \ref{nane_holmb}) obtained with the  
differents $\tau(R)$ we note that the time scale for the infalling gas affects the total number of dwarf stars  
produced by each model. The law presented by Renda et al. (2005) is the one which fits better the fraction of  
stars observed in the GCS of the solar neighbourhood, but the time scale for the solar radius that they adopted 
(see figure \ref{tau}), around $12$ Gyrs, is not realistic.  In this work the time scale for the infalling gas, was derived based on the distribution of dwarf stars in the solar neighbourhood, from which we know that it should be about 7 to 8 Gyrs assuming that the outermost regions of the Galaxy are still forming now. Also, this particular form of the $\tau(R)$ can fit the abundance, gas and SFR gradients, as we will show in the following sections.
 
On the other hand, the position of the peak and its wings are more important constraints,  
since owing to  
observational 
difficulties we still have problems to define the completeness of  
the survey (Holmberg et al. 2007). Focusing  
on these 
quantities one can note that time scales given by CMR2001 and this work reproduce very well the position of the peak in the observed distribution  
as well as the  
high metallicity wing,  whereas in the low metallicity side the number of stars is slightly overestimated, but this effect disappears  
if we consider other distributions (Figure \ref{nane_holmb}, right panel). Recently, Sch\"onrich \& Binney (2009) also 
reproduced the [Fe/H] distribution in the solar neighbourhood by means of a chemo-dynamical model suggesting that the G-dwarf 
distribution can be well explained by stellar migration, without considering a inside-out formation. However, the  
churning and blurring mechanisms invoked there imply a gas transfer that results in a similar effect.

After setting the ideal value for the time scale of the infalling gas ($\tau=0.75R+1.08$), a test with  
a variable star formation efficiency along the galactic disk was performed. An efficiency which decreases with the galactic  
radius was adopted until it 
 reached the lower-limit of $0.5 Gyr^{-1}$ at 12 kpc. This value is similar to that adopted in succesful models of dwarf irregular and spheroidal galaxies (see Lanfranchi \& Matteucci, 2003). In order to test our assumption of a variable $\nu$, we show in Figure 4 a plot of the empirical $\nu={SFR \over \Sigma_{gas}^{1.5}}$, obtained by adopting the observed SFR and  $ \Sigma_{gas}$ for the three galaxies. As one can see, for all the galaxies the ``observed efficiency'' shows a decreasing profile compatible with the 
trend used in our models.

\subsection{M31} 
 
M31 is a nearby spiral galaxy around two times more massive and 2.4 times larger than the Milky Way (Yin et al. 2009). It belongs to an earlier type than the Milky Way and possesses a larger bulge. 
 
To reproduce the chemical evolution of M31, we adopted the same model used for  
the Milky Way with the following modifications: 
 
\emph{Surface mass density distribution:} assumed to be exponential with the scale-length radius $R_D=5.4$ kpc and central surface  
density $\Sigma_0 = 460 M_{\odot}pc^{-2}$, as suggested by  Geehan et al. (2006). 
 
\emph{Time scale for the infalling gas:} $\tau(R)=0.62R+1.62$.  
This relation was derived under the assumption that at the galactocentric distance equivalent to the solar radius (the R corresponding to the $R_{\odot}$ calculated on the basis of  the $R/R_D$ ratio),  
M31 should have a time scale for the infalling gas similar to that of the solar vicinity. As well the outermost part of the optical  
disk 
is still accreting gas. 
 
\emph{Star formation efficiency:} the M31 present day gas profile in the disk shows a different trend relative to the Milky Way; it grows with the galactic radius and, 
after reaching a peak (at around 12 kpc) decreases steeply towards the center, thus suggesting a different scenario, relative to the Milky Way,  
for the star formation history of this galaxy. This trend is the signature of a very proeminent spiral arm  
detectable in 
M31 thanks to its inclination angle which allows us to measure the column density  of the hydrogen distribution. 
 
In order to reproduce the gas distribution we adopted three different star formation efficiencies. In model  
M31-A1 
we assumed $\nu=1 Gyr^{-1}$ (same as the Milky Way), in model M31-A2 it was set equal to $2 Gyr^{-1}$, whereas  
in model M31-B 
it was supposed to be a function of the galaxy radius $\nu(R) = 24/R - 1.5$, until it reached a minimum value  
of $0.5 Gyr^{-1}$ and then is 
assumed to be constant.

\emph{Threshold in the star formation:} we adopted a threshold in gas density of $5 M_{\odot}/pc^{2}$, as suggested in  
Braun et al. (2009). 
 
\emph{Exponent in the star formation law:} for this galaxy we also tested a model with a different exponent k in the
Schmidt law, using the lower limit given by Kennicutt et al. (1998), k=1.25 (this model is indicated by M31-Bk1.25).

\begin{figure}[!h]
\centering
\includegraphics[width=0.45\textwidth]{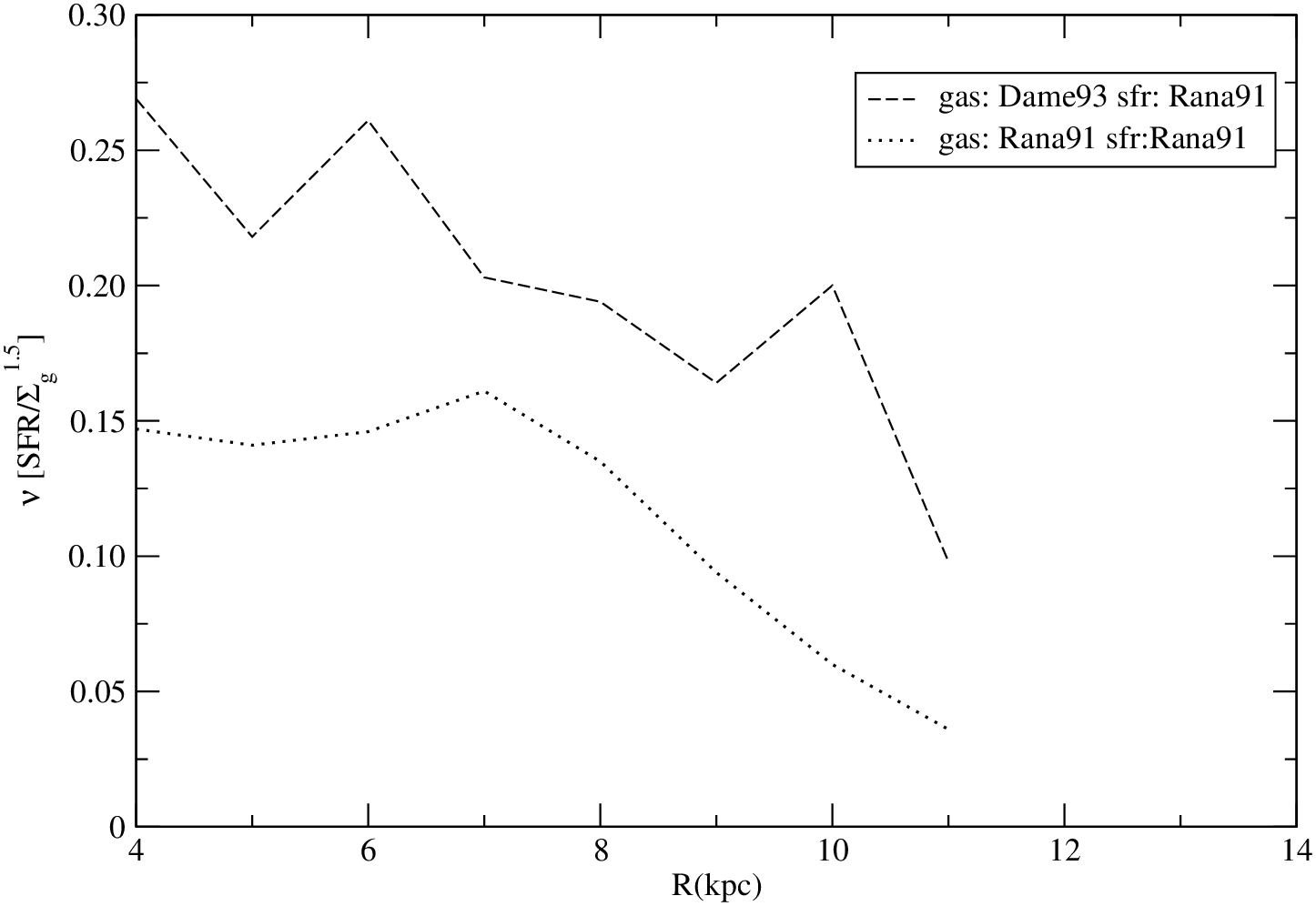}
\includegraphics[width=0.45\textwidth]{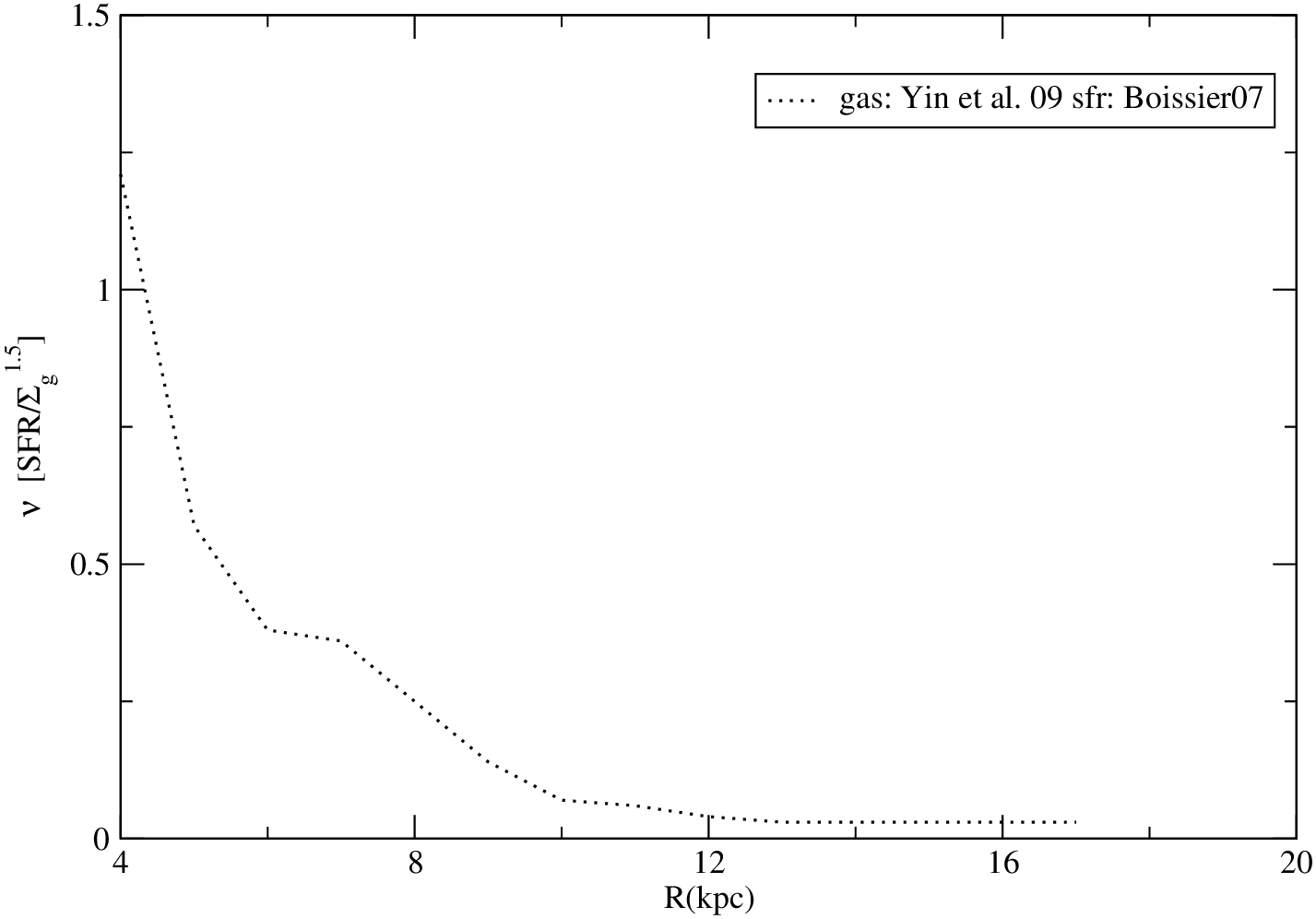}
\includegraphics[width=0.45\textwidth]{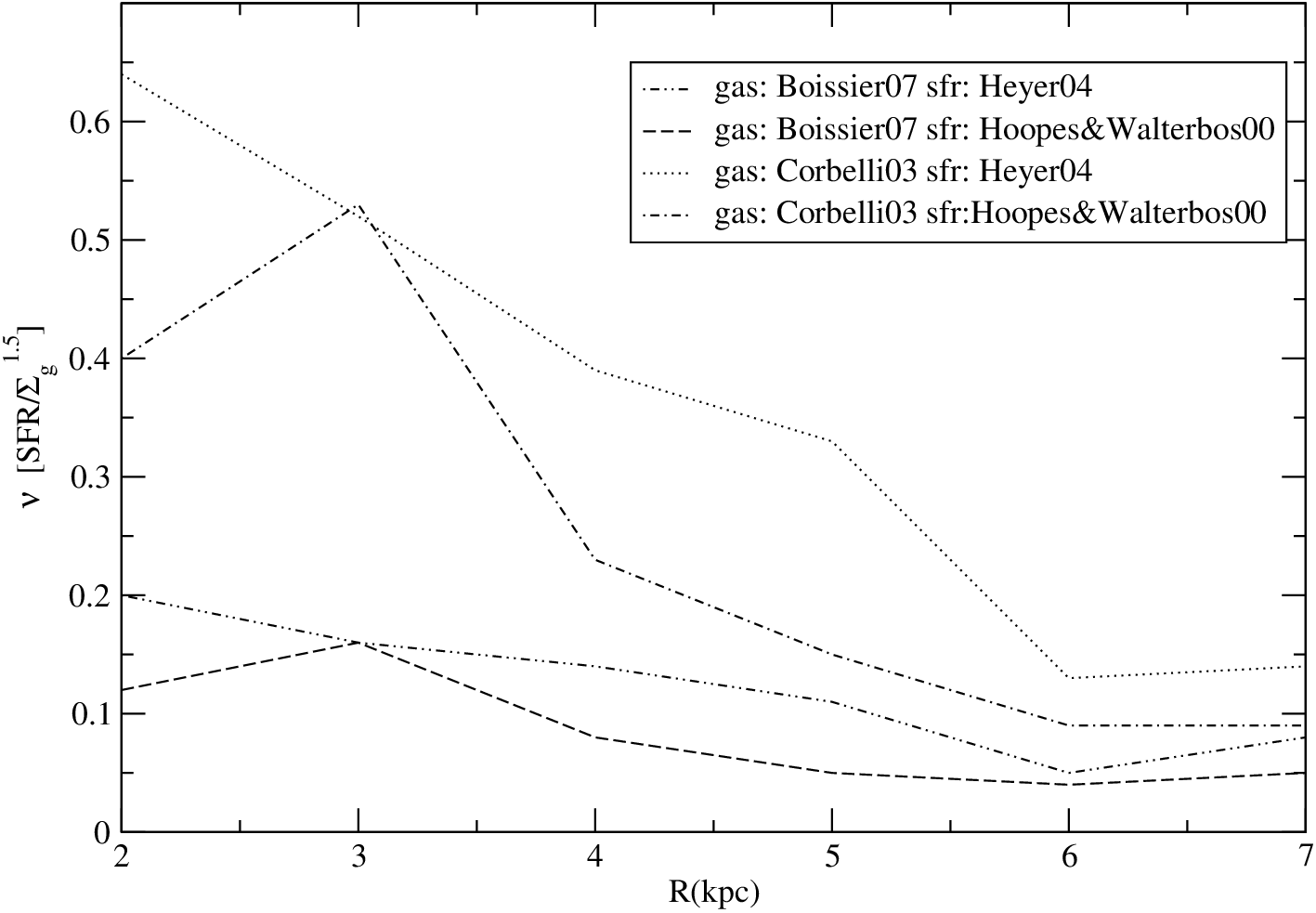}
\caption{ Estimate of the present day star formation efficiency using the observed gas surface density and
star formation rate. The first panel shows the values for  the MW (we used two different gas dsitributions and the SFR of 
Rana 1991), second panel is for M31 (using 
the gas distribution of Yin et al. 2009 and the SFR of Boissier et al. 2007)
while 
the last one shows the M33 values (using the gas distributions of Corbelli et al. 2003 and Boissier et al. 2007 and the SFR 
of Heyer et al. 2004 and Hoopes \& Walterbros 2000) .}
\label{nuR}
\end{figure}

\subsection{M33}

M33 or Triangulum Galaxy (NGC598) is a low density late type galaxy of the Local Group and has no remarkable sign of a bar or recent mergers.  
 
To model the chemical evolution of M33 we adopted the following parameters: 
 
\emph{Surface mass density distribution:} assumed to be exponential with the scale-length radius $R_D=2.2$ kpc (Corbelli 2003) and central surface density $\Sigma_0 = 230 M_{\odot}pc^{-2}$. 
 
\emph{Time scale for the infalling gas:} set in the same way as M31, as explained in section 2.2, with $\tau(R)=0.85R+4.54$  
 
\emph{Star formation efficiency:} for this galaxy we also used three different models. Two of them with constant efficiency equal to $\nu = 0.5 Gyr^{-1}$ and  
$\nu = 0.1Gyr^{-1}$ and another one with a efficiency that decreases with the radius: $\nu(R) = 1/R$, in this case we did not adopt a minimum value 
for the star formation efficiency because of the low density profile of this galaxy.

\emph{Threshold in the star formation:} $2 M_{\odot}/pc^{2}$. We adopted a smaller value for the threshold because of its diverse environmental conditions as a consequence of its lower mass.    
 
%\begin{figure*}[ht!] 
%%\sidecaption 
%\centering 
%\includegraphics[width=0.9\textwidth]{BW/grado_bw.eps} 
%\includegraphics[width=0.45\textwidth]{grado_01.eps} 
%\caption{Radial oxygen abundance gradient for the three galaxies in the sample. Milky Way data: HII Regions from Deharveng et al. (2000), Esteban et al. (2005), Rudolph et al. (2006), Cepheids from Andrievsky et al (2002a,b) and Planetary Nebulae from Costa et al. (2004). The left upper panes shows HII Regions observed in M31, data from: Galarza et al. (1999), Trundle et al. (2002) and Blair et al. (1982) + Dennefeld \& Kunth (1981). For M33 (left lower panel) the observed data are HII Regions from  
%Rosolowsky et al. (2007) and Type I Planetary Nebulae from Magrini et al. (2009). In these figures solid and dotted lines represent the models with constant $\nu$ while dashed 
%lines show the results for the models with variable $\nu$ (see table \ref{param} for details).  
%The right lower panel shows the models prediction with variable $\nu$ for all galaxies using  
%the normalised radius (solid line represents the MW, dashed M31 and dotted M33).} 
%\label{grado} 
%\end{figure*}  

\section{Results} 
 
In order to reproduce the observational chemical constraints of the spiral disks  
of three Local Group galaxies (MW, M31 and M33) and  to  
study the common features in the evolution of these systems, we computed several models by  
varying the three parameters shown in Table 2 ($\nu$, $\tau$ and the threshold).

\begin{table}[!ht] 
\caption{Models Parameters} 
\label{param} 
\begin{center} 
\begin{tabular}{ccccc} 
  \hline\hline 
  
\noalign{\smallskip} 
Model &  $\tau$ & $\nu$ & Thres. & Line \\ 
& $Gyr$ & $Gyr^{-1}$ & $M_{\odot}pc^{-2}$ & type \\ 
\noalign{\smallskip} 
 
  \hline 
\noalign{\smallskip} 
MW-A1 & 0.75R+1.08 & 1.0 & 7 & solid  \\ 
% &  &  \\ 
MW-A2 & 0.75R+1.08 & 1.0 & 4 & dots \\ 
MW-B & 0.75R+1.08 & 11/R-0.4 & 4 & dashed \\ 
% &  &  \\ 
M31-A1 & 0.62R+1.62 & 1.0 & 5 & solid \\ 
% &  &  \\ 
M31-A2 & 0.62R+1.62 & 2.0 & 5 & dots\\ 
% &  &  \\  
M31-B & 0.62R+1.62 & 24/R-1.5 & 5 & dashed \\ 

 M31-Bk1.25* &  0.62R+1.62 & 24/R-1.5 & 5 & dot-dashed \\
M33-A05 & 0.85R+4.54 & 0.5 & 2 & solid \\ 
M33-A01 & 0.85R+4.54 & 0.1 & 2 & dots \\ 
M33-B & 0.85R+4.54 & 1/R & 2 & dashed \\ 
 
\hline 
\end{tabular} 
\end{center} 
* For M31 we tested a model with a different exponent for the gas density in the SFR (k=1.25), using the lower value proposed by
Kennicutt (1998) $1.4 \pm 0.15$.
\end{table} 
 
In the following sections we compare our results for the radial gradient of oxygen abundance, present day gas and stellar distribution 
as well as SFR for all three galaxies. In each case a comparison between the disks is made using the normalized radius  
$R/R_D$. 
In figures \ref{grado}, \ref{gas} and \ref{sfr} solid and dotted lines represent the models with constant star formation efficiency 
$\nu$, namely MW-A1, MW-A2 M31-A1, M31-A2, M33-A05 and M33-A01. Dashed lines represent the models where the efficiency is a function  
of the galactic radius, $\nu(R)$, models: MW-B, M31-B, M31-Bk1.25 (dot-dashed) and M33-B (see table \ref{param} for details).  
In the bottom right  panel of those figures we present the comparison between the variable star formation efficiency models for each  
galaxy, where the solid line represents the Milky Way while dashed and dotted represent M31 e M33, respectively. 
 
\begin{table}[ht!] 
\caption{ Current values for the oxygen abundance gradient from the models for the galaxies in study} 
\label{valorigrad} 
\begin{center} 
\begin{tabular}{cccc} 
  \hline\hline 
  
\noalign{\smallskip} 
\multicolumn{4}{c}{\bf Milky Way}  \\ 
\noalign{\smallskip} 
  \hline 
\noalign{\smallskip} 
& MW-A1 & MW-A2 & MW-B  \\ 
\noalign{\smallskip} 
  \hline 
\noalign{\smallskip} 
Disk range & $\Delta$(O/H)/$\Delta$R & $\Delta$(O/H)/$\Delta$R & $\Delta$(O/H)/$\Delta$R \\ 
kpc & dex/kpc & dex/kpc & \\  
\noalign{\smallskip} 
  \hline 
4 to 14 &­ -0.059 ­& -0.025 & -0.029 \\ 
4 to 12 ­& -0.033 ­& -0.019 & -0.024 \\ 
4 to 10 ­& -0.024 ­& -0.017 & -0.026 \\ 
6 to 12 ­& -0.073 ­& -0.027 & -0.029 \\ 
\hline\hline 
 
\noalign{\smallskip} 
\multicolumn{4}{c} {\bf M31} \\ 
\noalign{\smallskip} 
  \hline 
\noalign{\smallskip} 
& M31-A1 & M31-A2 & M31-B (k=1.25)\\ 
\noalign{\smallskip} 
  \hline 
\noalign{\smallskip} 
Disk range & $\Delta$(O/H)/$\Delta$R & $\Delta$(O/H)/$\Delta$R& $\Delta$(O/H)/$\Delta$R\\ 
dex/kpc & dex/kpc & dex/kpc & dex/kpc\\  
\noalign{\smallskip} 
  \hline 
4 to 20 & ­-0.016 &­ -0.018 & -­0.020  (-0.031)\\ 
4 to 18 & ­-0.013 ­& -0.015 ­& -0.018  (-0.023) \\ 
4 to 16 & ­-0.011 ­& -0.014 ­& -0.019 (-0.017)\\ 
4 to 14 & ­-0.011 ­& -0.013 ­& -0.021 (-0.019)\\ 
6 to 16 & ­-0.010 ­& -0.012 ­& -0.017  (-0.015)\\ 
\hline\hline 
\noalign{\smallskip} 
\multicolumn{4}{c}{\bf M33}\\ 
\noalign{\smallskip} 
  \hline 
\noalign{\smallskip} 
& M33-A05 & M33-A01 & M33-B \\ 
\noalign{\smallskip} 
  \hline 
\noalign{\smallskip} 
Disk range & $\Delta$(O/H)/$\Delta$R & $\Delta$(O/H)/$\Delta$R & $\Delta$(O/H)/$\Delta$R \\ 
kpc & dex/kpc & dex/kpc & dex/kpc \\  
\noalign{\smallskip} 
  \hline 
2 to 6 & ­-0.015 & -0.015 & -0.063 \\ 
\hline\hline 
\noalign{\smallskip} 
\multicolumn{4}{c} {\bf Observed Values} \\ 
\noalign{\smallskip} 
  \hline 
\noalign{\smallskip} 
Galaxy & $\Delta(O/H)/\Delta R$ & References * & \\ 
 & dex/kpc& & \\ 
\noalign{\smallskip} 
  \hline 
 MW & -0.040 to -0.060 & 1, 2 & \\ 
 M31 & -0.013 to -0.027 & 3 & \\ 
 M33 & -0.012 to -0.054 & 4, 5, 6& \\ 
\hline\hline 
\end{tabular} 
\end{center} 
* References: (1)Deharveng et al. 2000, (2) Rudolph et al. 2006, (3) Trundle et al. 2002, (4) Crockett et al. 2006, (5) Magrini et al. 2007, (6) Rosolowsky et al. 2008. 
\end{table}

\subsection{Oxygen Abundance Gradient} 

\begin{figure*}[ht!] 
%\sidecaption 
\centering 
\includegraphics[width=0.9\textwidth]{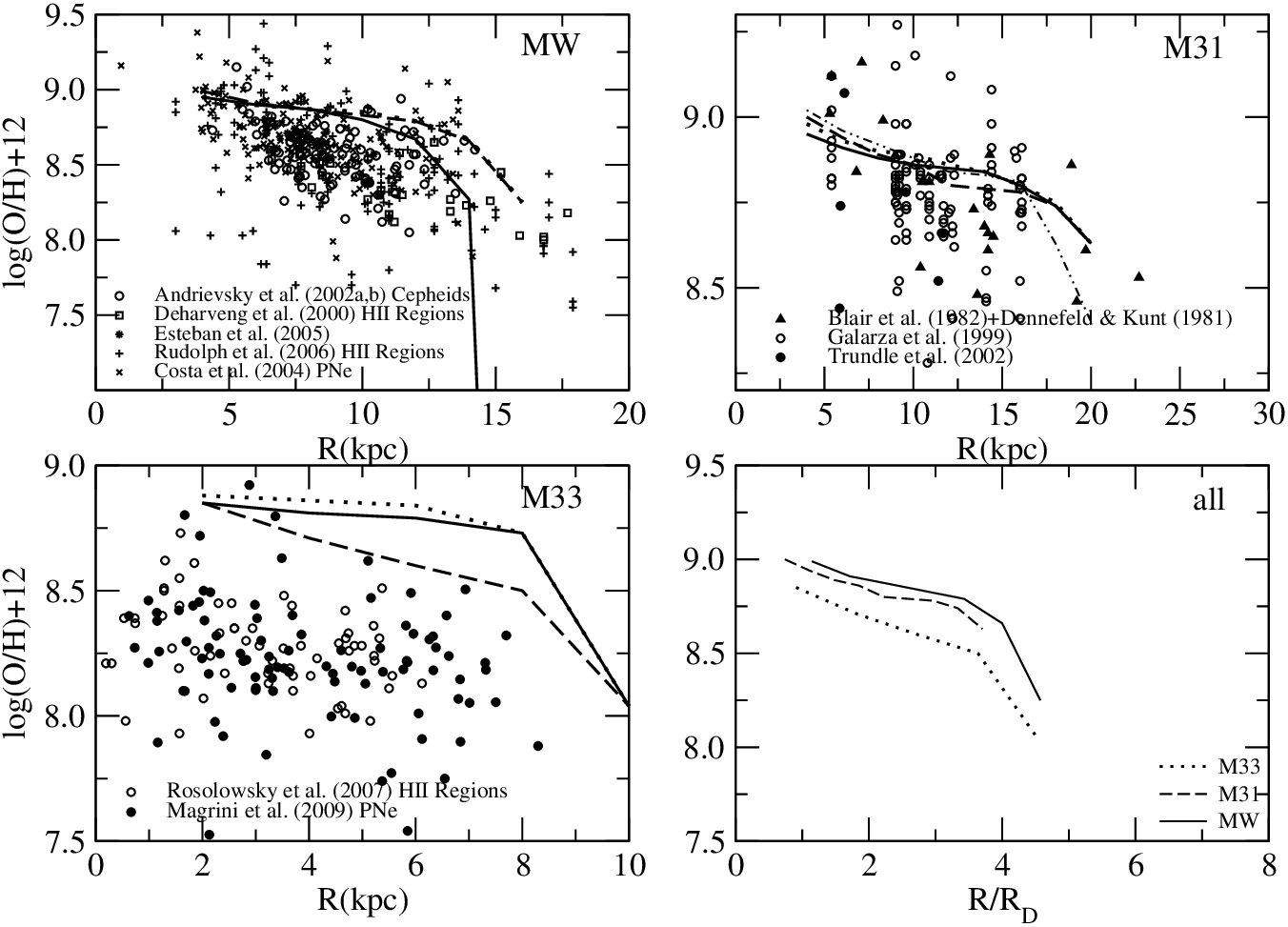} 
\caption{Radial oxygen abundance gradient for the three galaxies in the sample. Milky Way data: HII Regions from Deharveng et al. (2000), Esteban et al. (2005), Rudolph et al. (2006), Cepheids from Andrievsky et al (2002a,b) and Planetary Nebulae from Costa et al. (2004). The right  upper panes shows HII Regions observed in M31, data from: Galarza et al. (1999), Trundle et al. (2002), Blair et al. (1982) and Dennefeld \& Kunth (1981). For M33 (left lower panel) the observed data are HII Regions from  
Rosolowsky et al. (2007) and Type I Planetary Nebulae from Magrini et al. (2009). In these figures solid lines are for
MW-A1, M31-A1 and M33-A05 and dotted lines represent the models MW-A2, M31-A2 and M33-A01 while dashed 
lines show the results of the models Mw-B, M31-B (dot-dahed for M31-Bk1.25) and M33-B (see table \ref{param} for details).  
The right lower panel shows the models prediction with variable $\nu$ for all galaxies using  
the normalised radius (solid line represents the MW, dashed M31 and dotted M33).} 
\label{grado} 
\end{figure*}  
 
The effect of the variable efficiency in the SFR can also be noted in our models. As is expected, when comparing models with the 
same threshold we can see that those  with $\nu(R)$ present a steeper gradient than the models with constant 
efficiency.

In figure \ref{grado} we show the results for the oxygen abundance gradient for the Milky Way (left upper panel), M31 (right upper panel), 
M33 (left lower panel) with a compilation of observational data and a comparative plot between the predictions for the galaxies (right lower panel).  
The abundance results are slightly shifted for the MW and overestimated for M33, but the predicted slopes of the gradients are in agreement with the observational data. The threshold effect can be seen in the breaks in the gradients being more remarkable for the MW and M33 as M31 (except when k=1.25) has a more massive disk. 
Table \ref{valorigrad} shows the present day values of the radial abundance gradient of the oxygen in different galactocentric distance ranges as predicted from our models compared with the observed values for the three galaxies.  
The predictions of all our models are in very good agreement with observational data and the effect of different  
thresholds can be noticed in the results of the Milky Way models with constant efficiency in the SFR. Model MW-A1 
 (threshold of $7 M_{\odot}pc^{-2}$) presents gradients much steeper than the values predicted by MW-A2  
(threshold equal to $4 M_{\odot}pc^{-2}$) which is a consequence of the suppression of the star formation.  

\begin{figure*}[ht!] 
%\centering 
\includegraphics[width=0.9\textwidth]{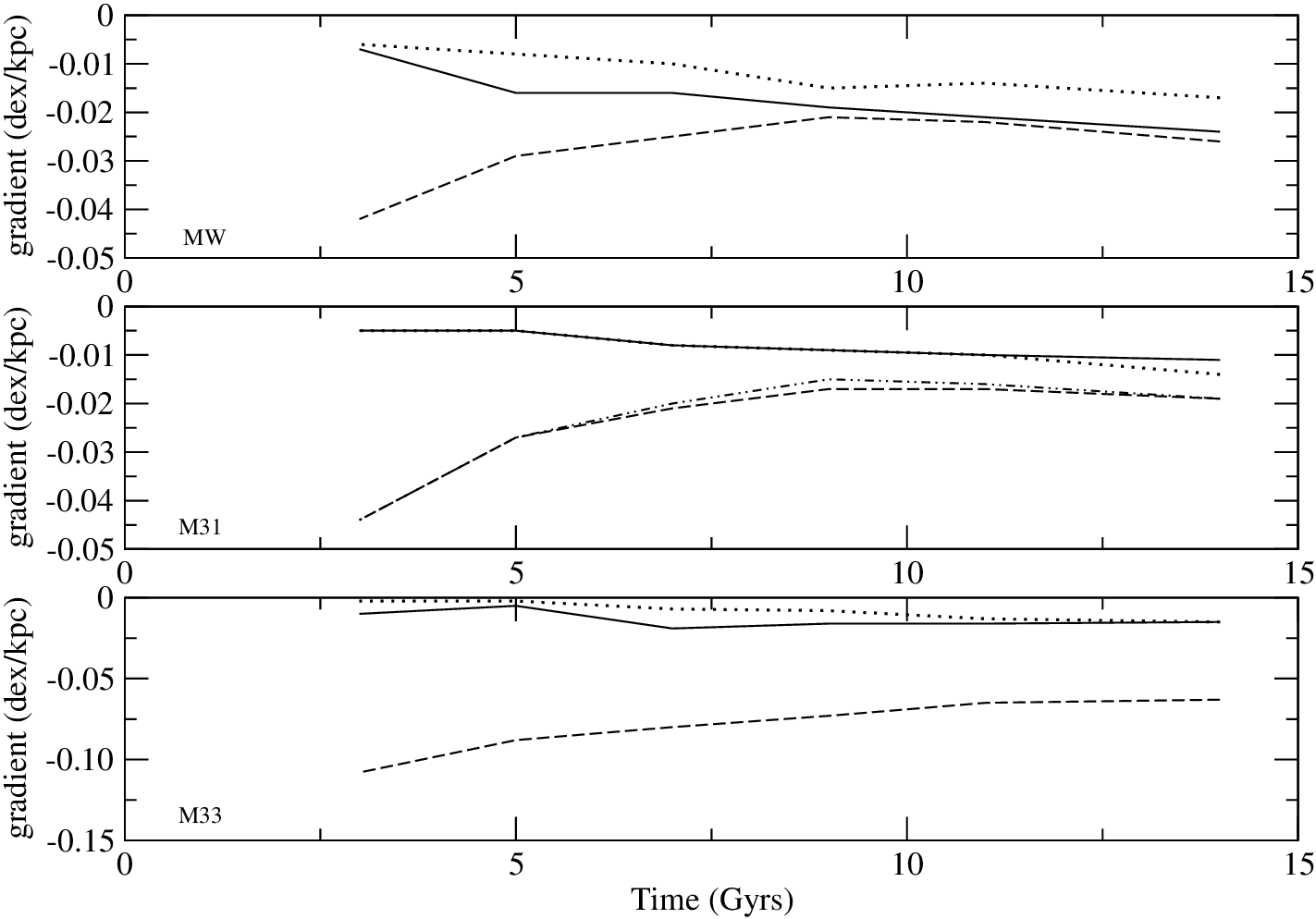} 
\caption{Time evolution of the slope of the radial abundance gradient of oxygen for all models. The first panel shows  
the results for the Milky Way (solid line for Mw-A1, dotted line for MW-A2 and dashed line for MW-B), while
the middle panel is for M31 (solid line for M31-A1, dotted line for M31-A2, dashed line for M31-B and dot-dashed line for
M31-Bk1.25) and the last panel shows the predictions for M33 (solid line for M33-A05, dotted line for M33-A01 and dashed line for M33-B).
See table 2 for more details.  } 
\label{evolgrad} 
\end{figure*}

%%% Retirar figuras da evolução do gradiente %%%%%%%%%%%%% 
% 
%\begin{figure*}[ht!] 
%\label{gradopne} 
%\centering 
%\includegraphics[width=0.9\textwidth]{gradpne_MW3g.eps} 
%\includegraphics[width=0.45\textwidth]{M31grado_pne.eps} 
%\caption{Radial gradient of oxygen abundance for the Milky Way (panels 1, 2 and 3) at different times.Solid line represents the present day gradient, while 
%dashed and dotted lines show the values of the gradient at 7 and 3 Gyr, respectively. Open circles represents planetary 
%nebulae from Costa et al. 2004}  
%\end{figure*} 
% 
% 
%\begin{figure*}[ht!] 
%\includegraphics[width=0.45\textwidth]{M31grado_pne.eps} 
%\includegraphics[width=0.45\textwidth]{m33grado_evolnew.eps} 
%\caption{Radial gradient of oxygen abundance for M31 (left panels) and M33 
%(right panels) at  different times. Solid line represent the present  
%day gradient, while dashed and dotted lines show the values of the gradient  
%at 7 and 3 Gyr, respectively. M33 planetary nebulae data are from Magrini et al. 2009} 
%\label{gradopne2} 
%\end{figure*}  

Figure \ref{evolgrad} shows the evolution of the oxygen abundance gradient with time. Models are the same  as those presented in  
table \ref{param}. 
 
It is  
not easy to establish if the radial abundance gradient tends to flatten or steepen with time;  
models can predict flattening or steepening of the gradient depending on the adopted SFR along the disk.  
As one can see, for $\nu(R)$ all models predict a gradient which flattens with time.  
Gradients which flatten out  
in time with a decreasing flattening rate in the last few Gyrs are supported by models such as those proposed  
by Hou et al. (2000), Moll\`a \& Diaz (2005) and Magrini et al. (2009), while models proposed by Tosi (1988) and CMD2001  
predict the steepening of the gradients in time. Observational results by Maciel et al. (2003) support the  
flattening in time of the oxygen abundance gradient in  
the Milky Way,  while Magrini et al. (2009) also noted a flattening in the oxygen gradient in their sample of M33  
planetary nebulae.   
 
%\begin{figure}[ht!] 
%\centering 
%\includegraphics[width=0.45\textwidth]{BW/evol_grad_bw.eps} 
%\caption{Time evolution of the radial abundance gradient of oxygen for all models. Solid lines and dotted  
%lines represent the models with constant efficiency and dashed lines represent models with variable efficiency. } 
%\label{evolgrad} 
%\end{figure}  

%\begin{figure}[ht!] 
%\includegraphics[width=0.45\textwidth]{stars_01.eps} 
%\caption{Comparative plot of stellar surface density for all galaxies in  
%function of Radius in kpc. Black lines represent 
%the MW models, red lines are the M31 models and green lines presents the results for M33. The shaded area corresponds to 
%observed values for the Milky Way.} 
%\label{stars} 
%\end{figure}  

%\begin{figure} 
%\centering 
%\includegraphics[width=0.45\textwidth]{pn_m33grado.eps} 
%\caption{Radial oxygen abundance gradient for M33 in different times. } 
%\label{gradopnem33} 
%\end{figure} 

\subsection{Gas and stellar distribution}

\begin{figure*}[ht!] 
%\sidecaption 
%\centering 
\includegraphics[width=0.9\textwidth]{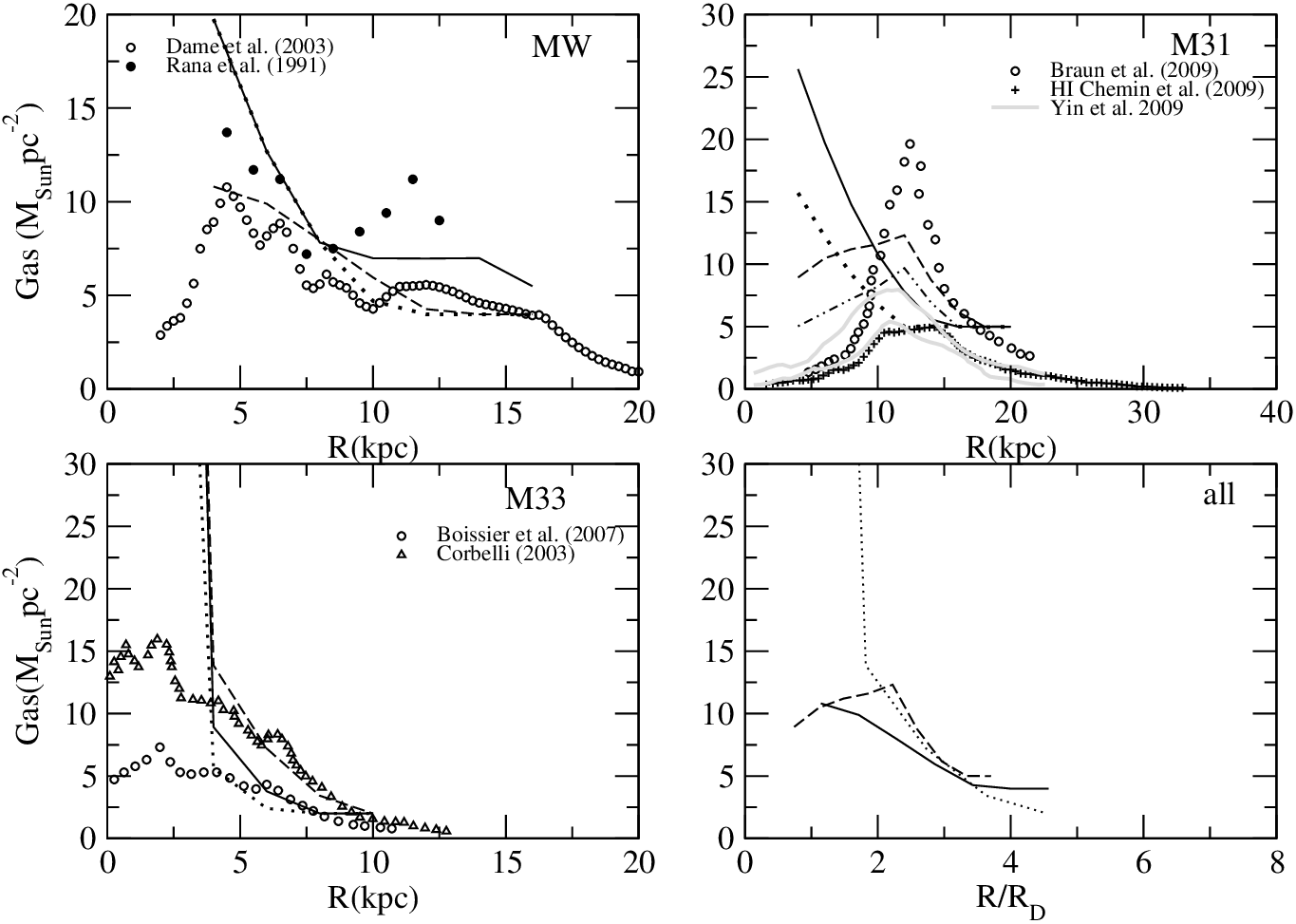} 
\caption{Present-day radial gas distribution. Milky Way data: Rana (1991) and Dame et al. (1993).  M31: solid circles represents the
total gas data from Braun et al. (2009) and crosses represents HI data from Chemin et al. (2009). M33: data from Boissier et al. (2007) and Verley et al (2008). 
In these figures solid lines are for
MW-A1, M31-A1 and M33-A05 and dotted lines represent the models MW-A2, M31-A2 and M33-A01 while dashed 
lines show the results of the models Mw-B, M31-B and M33-B (see table \ref{param} for details).
The right lower panel shows the model prediction with variable $\nu$ for all galaxies using  
the normalised radius (solid line represents the MW, dotted M33 and dashed M31).}
\label{gas} 
\end{figure*}

Figure \ref{gas} shows the present day gas surface density distribution for the galaxies studied here. For the Milky Way we note that the model MW-B describes better the present surface gas density of the disk with a smoother distribution while MW - A1 and MW - A2 
present a remarkable break in the profile (probably associated with the constant efficiency in the SFR).  
 
For M31 the models with constant star formation efficiency can reproduce an exponential profile but fail to explain the peak 
located at about 12kpc of distance from the galaxy center. The models with variable efficiency (M31-B and M31-Bk1.25) can instead
reproduce this peak as shown in figure \ref{gas}.
 
Both models for M33 are capable of explaining the present day gas distribution of Boissier et al. (2007) for R$>$ 4 kpc 
but overestimate the gas surface density in the inner regions.

%\begin{figure}[ht!] 
%\centering 
%\includegraphics[width=0.45\textwidth]{BW/stars_bw.eps} 
%\caption{Comparative plot of stellar surface density for all galaxies in  
%function of Radius in kpc. Black lines represent 
%the MW models, red lines are the M31 models and green lines presents the results for M33. The shaded area corresponds to 
%observed values for the Milky Way} 
%\label{stars} 
%\end{figure}  

In Figure \ref{stars}, we show the predicted stellar density distributions along the three disks. The Milky Way models  
are in good agreement with the observed values (shaded area). All this area was obtained by assuming an exponential 
distribution (see also CMR2001) and using a scale length radius 
$R_{Dstars}$ equal to 2.5 kpc (Freudenreich 1998) together with a stellar density for the solar annulus equal to $35 M_{\odot}pc^{-2}$ (Gilmore et al. 1998), in order to scale the distribution to the observed values
. The predictions for M31 stellar density profile  
show a shallower distribution, whereas for M33 our models present a steeper distribution (even without taking into account the inner 
part of the M33 disk which is also overestimated as a consequence of the gas profile - see figure \ref{gas}).  
   
\subsection{SFR} 
 
\begin{figure*}[ht!]
%\sidecaption
%\centering
\includegraphics[width=0.9\textwidth]{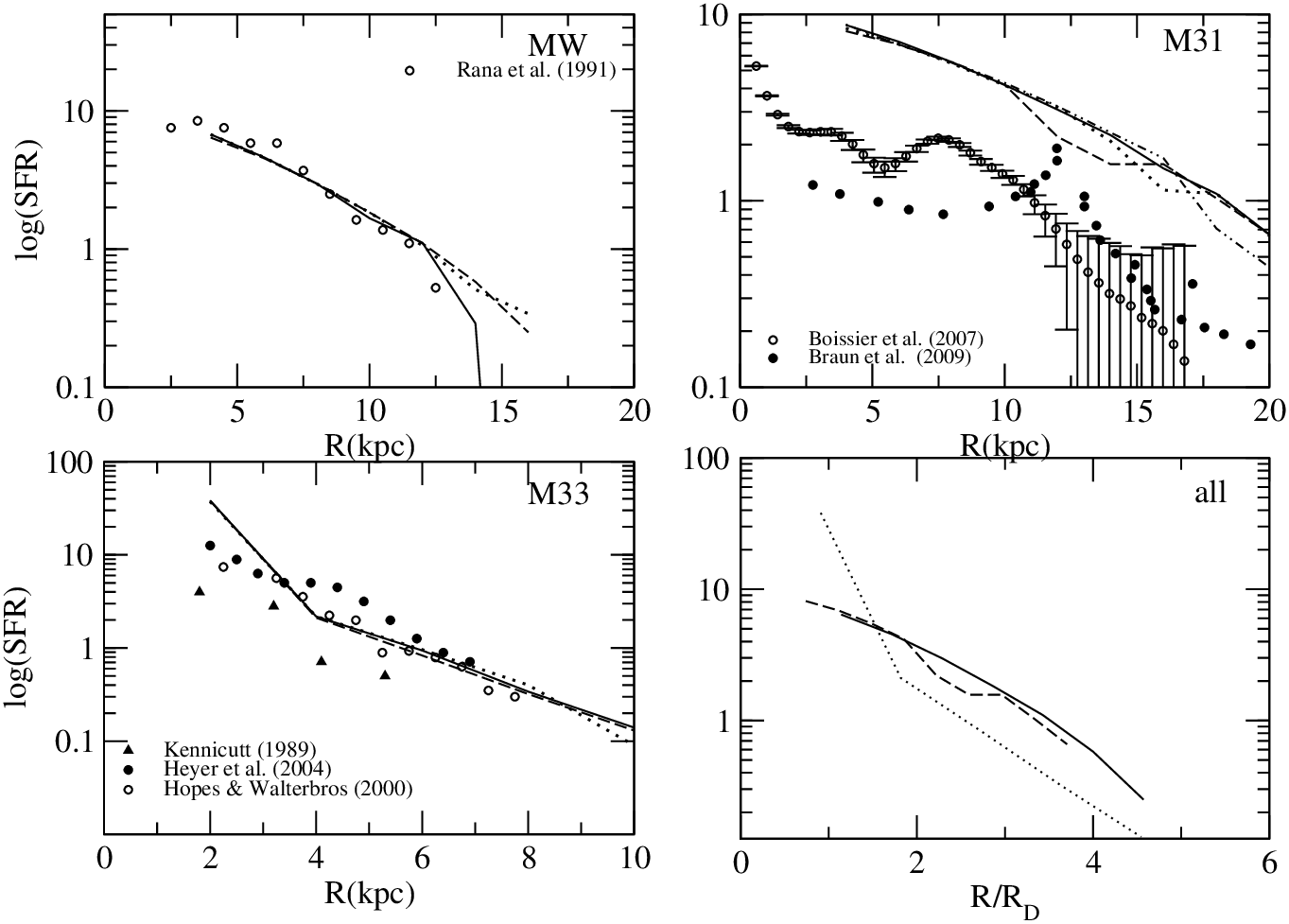}
\caption{Present-day radial distribution of the SFR. Observational data for the Milky Way are those from Rana et al. 1991, for M31 open circles represents the data from
Braun et al. 2009 and for M33 the observation data are from Hoopes \& Walterbros (2000), Heyer et al. (2004) and Kennicutt (1989).
In these figures solid lines are for
MW-A1, M31-A1 and M33-A05 and dotted lines represent the models MW-A2, M31-A2 and M33-A01 while dashed 
lines show the results of the models Mw-B, M31-B, M31-Bk1.25 (dot-dashed) and M33-B (see table \ref{param} for details).
The right lower panel shows the models prediction with variable $\nu$ for all galaxies using
the normalised radius (solid line represents the MW, dotted for M33 and dashed for M31).}
\label{sfr}
\end{figure*}

\begin{figure}[ht!] 
%\centering 
\includegraphics[width=0.45\textwidth]{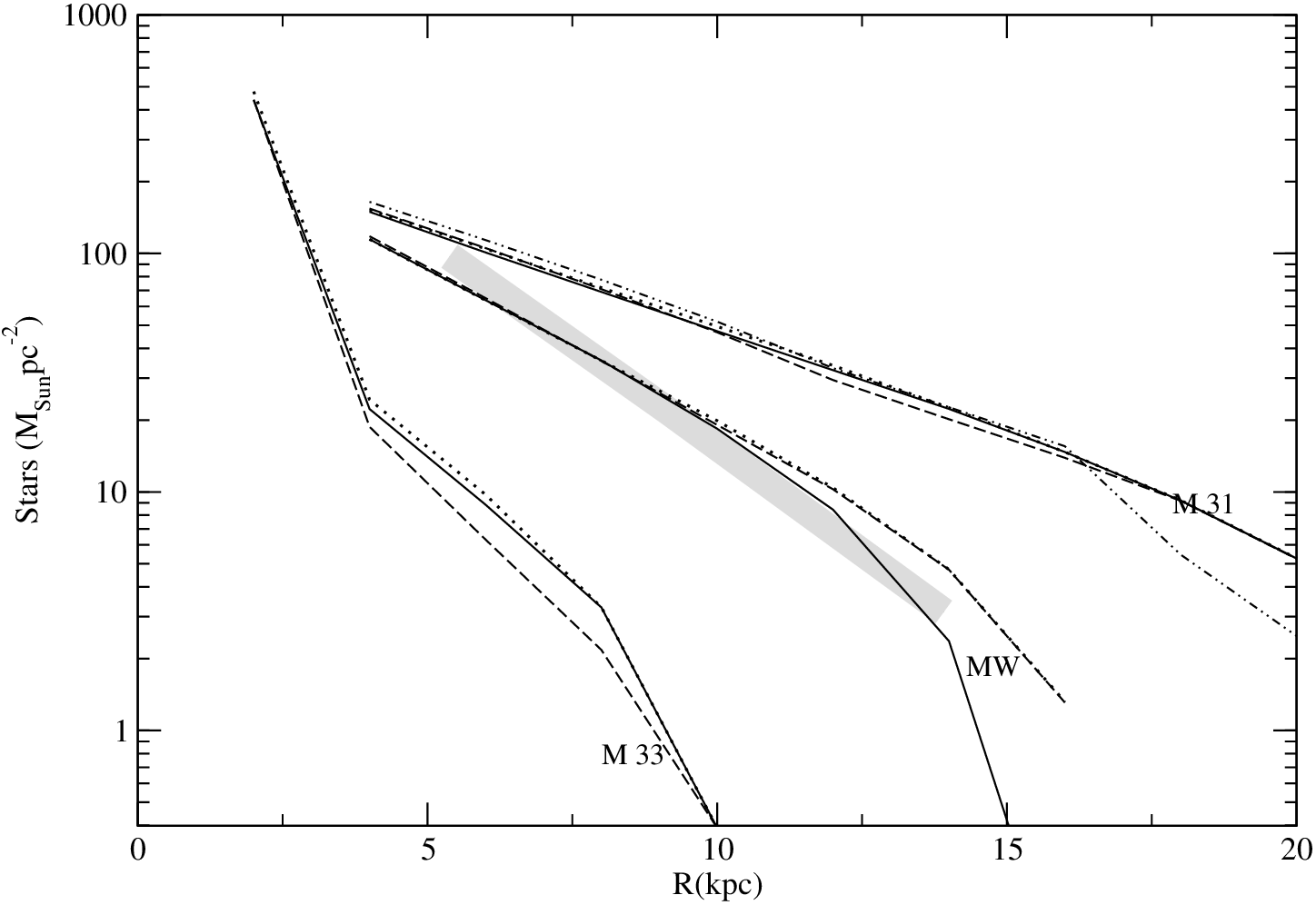} 
\caption{Comparative plot of stellar surface density for all galaxies in  
function of Radius in kpc. For the MW solid line for MW-A1, dotted line for MW-A2 and dashed line for MW-B,
for M31 solid line represents M31-A1, dotted line M31-A2, dashed line M31-B and  dot-dashed line for M31-Bk1.25, finally for M33 
solid line is for M33-A05, dotted line for M33-01 and dashed line for M33-B.
See table 2 for more details. The shaded area corresponds to a scaled exponential distribution with $R_{Dstars} = 2.5$kpc (Freudenreich 
1998) using the local values for the stellar density ($35 M_{\odot}pc^{-2}$, Gilmore et al. 1989).} 
\label{stars} 
\end{figure}

In Figure \ref{sfr} we show the predicted distributions of the SFR in the three disks compared to observational data. All the models can reproduce 
observational trends and the differences between the predictions of different models for each galaxy is very small. For the MW, the models differ  
only in the outer region where the threshold in the star formation it can be noticed that the model MW-A1 (with the highest
threshold density value) present a steeper profile. The predictions for M31 show a higher SF than the observed data and this fact could be related to the the limitations 
of the method used to estimate the SFR in M31 (uncertanties in the adopted IMF and assumed metallicity in the conversion
factors from the UV to the star formation rate values)
The models M33-A05 and M33-B have similar results for the star formation rate in very good agreement with the observational data  
(except for the innermost region).

\subsection{Deuterium Abundances} 

\begin{figure}[ht!] 
%\centering 
\includegraphics[width=0.45\textwidth]{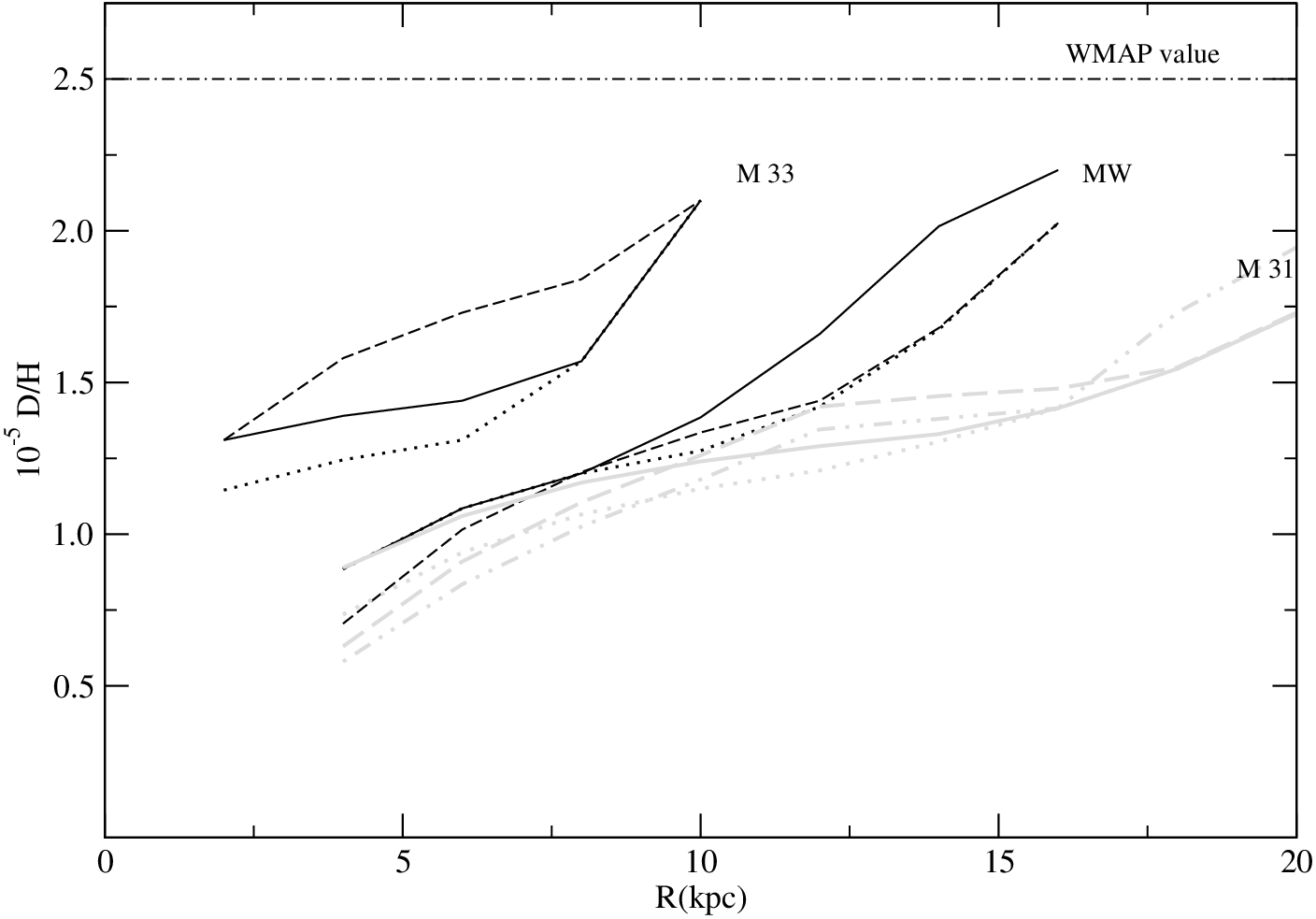} 
\caption{Deuterium gradient for all galaxies in function of radius in kpc.  
For the MW solid line for MW-A1, dotted line for MW-A2 and dashed line for MW-B,
for M31 solid grey line represents M31-A1, dotted grey line M31-A2, dashed grey line M31-B and dot-dashed grey line for
M31-Bk1.25,  finally for M33
solid line is for M33-A05, dotted line for M33-01 and dashed line for M33-B. See table 2 for more
details.
} 
\label{deut} 
\end{figure}

Figure \ref{deut} shows the deuterium radial abundance gradient for the three galaxies studied in this work. Previous studies have already shown that the abundance of D/H in the disk of the MW should increase with radius (e.g. Prantzos, 1996; Romano et al. 2006), but this is the first time that this gradient is computed for external galaxies. One can note that the deuterium gradient has the opposite behaviour to the oxygen distribution along the galactic radius, reflecting the fact that D is only destroyed during galactic evolution. We show this diagram just as a prediction, since there are no data for M31 and M33 and for the MW there are data only for the local ISM. 

In particular, the measures of the local abundance of D  show a large spread indicating different values along different lines of sight. The most plausible interpretation for this is that D can condense into carbon grains and PAH molecules, thus being removed from the ISM (Linsky et al. 2006 and references therein).  Moreover, there is another problem and that is that the largest D abundance measured locally is higher than expected from chemical evolution models, thus implying an astration factor of $1.12 \pm 0.13$ (Savage et al. 2007), while our model, for example, predicts a factor of $\sim 1.5$ (we assume a primordial $(D/H)_p= 2.5\times10^{-5}$ in number, as suggested by WMAP). Some additional information on the primorddial D/H abundance comes from the measures in QSO absorbers which provide a lower limit for the primordial D/H abundance. These systems, in fact, are found at high redshift $z>2$ and their metallicity is low and therefore the D/H, which is only destroyed during galactic evolution, is likey to be close to the primordial value. In particular, Pettini et al. (2008) by analysing several QSO absorbers, where the most distant is at $z=2.61843$ and has an oxygen abundance $\sim 1/250 O_{\odot}$, conclude that the estimated average primordial D value is $log<(D/H)>_p=-4.55 \pm 0.03$, which is very close to the value determined by WMAP.
It is difficult to reconcile chemical evolution with such a low D astration, so the question is still open. Since D is only destroyed during galactic chemical evolution, an infall of gas with primordial chemical composition could help in raising the D abundance. However, our model already well fits the G-dwarf metallicity distribution with gas infall of almost primordial chemical composition.
In this context, we only want to show how different histories of star formation in different galaxies produce different D gradients along the disk.The D astration factor predicted for M33 is $1.2$ and for M31 is $1.6$,  reflecting the lower and higher SFR of these two galaxies, respectively relative to the MW.
 
\section{Discussion} 
 
In this work we used a one-infall generalised model to reproduce the chemical evolution of the disks of spiral galaxies in the Local Group.  
We focused this study on the effects of different star formation efficiencies ($\nu$), in the chemical evolution of disks.  The main differences  between models of different galaxies are the efficiency of star formation and the timescales for the disk formation at different radii (inside-out process).

Concerning the star formation threshold effect in the radial oxygen gradients we note that it is more  
visible for the Milky Way and M33 than for  M31 (except for k=1.25). This fact is compatible with the scenario proposed by Pohlen et al. (2004) who 
suggest that the star formation threshold can produce a truncation in the observed stellar luminosity profile  
of spiral disks and that low-mass 
galaxies should have smaller values for this radius than the massive ones. 

The present-day chemical abundance gradients for oxygen, as predicted by the models, are in good agreement  
with observational data and can assume 
different values depending on the star formation efficiency  and threshold in the  
surface gas density used in the models. The time evolution of the  
gradients also reflects this trend as  
the results show that it can steepen or flatten in time depending on the adopted value of $\nu$. For the Milky Way,  
models MW-A1 and MW-A2 present an oxygen gradient which steepens in time like in CMR2001, which was already expected since this study is based on an updated version of that model, but model MW-B  
shows a different behavior which confirms the relation between the star formation efficiency and  
the evolution of the radial abundance  
gradients.  
The same happens to M31 where the oxygen abundance gradient flattens during the galaxy evolution if a constant efficiency
is adopted and shows an steepening when a variable efficiency is used.
 
For M33, the model results show a similar trend for the evolution of the gradient but with different absolute values,  
reflecting the diverse star formation history induced by the different efficiencies. 
Renda et al. (2005) and Yin et al. (2009) have also modeled the chemical evolution of the Milky Way and M31. In their model, Renda et al. (2005) used a star formation law exponent  
of $k=2$ for the disk at variance with Kennicutt (1998) law,  whereas Yin et al. (2009) used a star formation efficiency higher than the one used in the MW and variable with galactic radius. Our results show that for the MW the model 
with variable $\nu$ and threshold equal to $4M_{\odot}/pc^2$ produces a very good agreement with the observational data and that for M31 the model with $\nu(R)$ can reproduce the peak around 12 kpc of the present day total gas surface distribution. It is interesting 
to note that when we keep all the parameters fixed and change only the exponent of the SFR  k=1.25, the gas distribution goes down to lower
values and gets closer to the upper limit of the Yin et al. (2009) data. This implies that the exponent of the SF law has a strong effect on the gas distribution.

 The models for M33 produce results in good agreement with the observational data in the outer region of the galaxy but fail to reproduce the present day gas content in the inner region of M33 disk. Magrini et al. (2007) also predicted an overestimated gas content in the inner kpcs of M33 disk. Perhaps the lower than predicted gas content could be attributed to some bulge-disk interaction effect. 
 
The present-day stellar surface mass density of the Milky Way is in good agreement with observational data and the threshold effect can be clearly seen 
in the outer disk of the Galaxy, where model MWA-1 (with the highest threshold value) shows a steeper behavior demonstrating that the star formation has been suppressed. Comparing the distributions of the stellar density in all galaxies we note that it gets flatter going  from M33 to M31, thus indicating a possible relation between the galaxy total surface mass density and the slope of the stellar distribution.

\section{Conclusions}

%\begin{itemize} 
\subsection{The MW} 
%\item  
%{\bf MW} 
 
We found that the oxygen gradient along the disk of the Milky Way is well reproduced if an inside-out disk 
formation is assumed together with a threshold in the star formation of $7M_{\odot}pc^{-2}$ or $4M_{\odot}pc^{-2}$ ,  
in agreement with previous works (CMR2001, Colavitti et al. 2008). The present time radial oxygen gradient is very dependent 
on the threshold in the star formation while it seems not to be so sensitive to the efficiency ($\nu$) of the SFR.  
 
The oxygen gradient can either flatten or steepen in time according to the assumption made on the star formation efficiency as a function of galactocentric distance. 
Models with a constant $\nu$ tend to predict a steepening of the gradients in time, whereas those with a $\nu$ decreasing with  
the radius  tend to flatten (in agreement with some recent observations of Maciel et al. 2003) 
Clearly the gradient evolution with time is strongly related to the assumed history of star formation in the disk. 
 
The present-day gas profile in the MW is better reproduced by the model with a threshold of $4M_{\odot}pc^{-2}$ and $\nu(R)$. 
All models predict a lower SFR for the inner disk of the Galaxy but are in very good agreement with the observed data  
in the solar neighbourhood and in the outer parts of the disk. The higher SFR in the inner parts of the MW disk can be due to  
the presence of a bar, as suggested in Portinari \& Chiosi (2000), and therefore cannot be reproduced by simple chemical evolution models. 
 
The stellar surface density of the MW is in agreement with the observed values but models with a smaller threshold (MW-A2 and 
MW-B) overestimate the stellar content in the outer region of the disk. Unlike other observational constraints, the 
variable efficiency in the SFR does not play an important role for the results relative to the stellar sufarce density, indicating that the threshold is a stronger mechanism to regulate it.  
 
In summary, the model that fits best the observational constraints for the Milky Way is the model with variable efficiency for the the star formation (MW-B).

\subsection {M31} 
 
The evolution of the disk of M31 is well reproduced by assuming a faster evolution (faster means a more intense SFR which is due both to the higher efficiency of SF and to the shorter infall timescale) than in the disk of the MW and a higher star formation threshold. 
Since the disk of M31 is more massive than the MW one, this implies that more massive disks 
should form faster and therefore that they are older than less massive ones (see also Boissier et al. 2003). 
The O abundance gradient from HII regions is well reproduced by all models for M31. This result confirms the predictions for the MW, showing that the present day abundance gradient is not so sensitive to the changes in the star formation efficiency. On the other hand, the time evolution of the O gradient is very dependent on this efficiency, steepening or flattening in time according to the chosen $\nu$.  This fact can be confirmed with the results obtained for M31-Bk1.25 where we used the same efficiency but a different 
exponent for the SFR.

Models with constant efficiency in the star formation (M31-A1 and M31-A2) provide an exponential distribution of the present day gas surface density,  while models M31-B and M31-Bk1.25 with variable efficiency predict a more realistic scenario with a peak in the gas distribution around 12 kpc which can be related to M31 spiral arms. 
 
The stellar density profile is flatter than the one predicted for the MW and M33 and all models show a similar distribution. 
 
The predicted SFR for M31 is very similar in all models, specially M31-A2 and M31-B that present a smaller star formation
after the peak in 12 kpc 

In summary, the best model for the M31 disk is also the one with a star formation efficiency varing through the  
disk with a lower exponent in the SF law (M31-Bk1.25).
 
%\item  
%{\bf M33} 
\subsection{M33} 
 
The chemical evolution of the M33 disk is reproduced with a slower evolution and lower star formation threshold than in the MW and M31  
 
The slope in the abundance gradient is well reproduced, but the oxygen abundances are overestimated by 0.25 dex. This is an 
indication that the chemical evolution models used for large spiral galaxies need to be adjusted to reproduce the 
abundances of smaller and lower massive disks.  In any case, the time evolution of the abundance gradient is also very  
dependent on the chosen efficiency in the SFR, as can be seen for the MW and M31. 
 
The models fail to reproduce the present day gas profile in the inner disk, whereas it is very well reproduced for $R> 5$ kpc 
(same problem faced by Magrini et al. 2007), indicating a possible bulge-disk interaction in this region despite   
the small visible bulge of M33.  
 
Compared to the other galaxies in this sample,  M33 presents the steeper stellar distribution along the radius, without 
taking into account the inner regions of the galaxy. 
 
The SFR predicted for  M33 is in very good agreement with observations and is the one that is best reproduced by our models. 
 
For M33, our best model is the one with the lower efficiency, constant along the galaxy radiusi (M33-A01). This fact suggests that the 
star formation history in small and low density disks is probably different from the more massive ones such as the MW and M31.
 
\par  
 
In conclusion we find that the present day value of the oxygen abundance is more sensitive to the threshold in the SFR than to 
the efficiency in the star formation and that this latter parameter plays an important role in the time evolution of the 
gradient. The variable efficiency in the SFR is also important to reproduce the present day gas distribution in the disk 
of  galaxies with a marked presence of spiral arms. A correlation between the galaxy mass and the star density 
profile can be seen when observing that the stellar distribution along the galactic radius gets steeper from the most  
massive (M31) to the lower massive one (M33). Another interesting result is the dependence of the gas distribution along the disks of spirals on the exponent of the Kennicutt law. By varying this exponent by $\pm0.15$, which corresponds to the observational error,  one can obtain very different gas distributions.

{\it An important conclusion of this paper is that there should be a downsizing in star formation also in spirals, similar to what applies to ellipticals. A similar conclusion was reached by Boissier et al. (2003).}

%\end{itemize} 
 
%\section*{Acknowledgments} 
\begin{acknowledgements} 
We thank F. Calura, G. Cescutti, E. Spitoni, S. Scarano Jr., E. M. Rangel and I. J. Danziger for many useful discussions. We acknowledge financial support from the 
CNPq (Processes: 302538/2007-0 and 200412/2008-6), FAPESP (2006/59453-0), MIUR PRIN2007, Prot.2007JJC53X-001.  We also would like to
thank the referee S. Boissier for his constructive suggestions.  
 
\end{acknowledgements}

\end{document}